\documentclass[aps,floatfix,nofootinbib,twocolumn,prl]{revtex4-1}

\newcommand{\eq}[1]{(\ref{#1})}
\newcommand{\be}{\begin{equation}}
\newcommand{\ee}{\end{equation}}
\newcommand{\bea}{\begin{eqnarray}}
\newcommand{\eea}{\end{eqnarray}}

\newcommand{\vs}[1]{\vspace{#1 mm}}
\newcommand{\hs}[1]{\hspace{#1 mm}}

\def\a{\alpha}
\def\b{\beta}

\def\d{\delta}

\def\e{\epsilon}

\def\f{\phi}
\def\fr{\frac}

\def\m{\mu}
\def\n{\nu}

\def\s{\sigma}

\def\th{\theta}

\let\bm=\bibitem
\def\nn{\nonumber}

\begin{document}
\large

\title{A Simple Proof of Locality in Quantum Mechanics}

\author{Ali Kaya}

\email[]{alikaya@tamu.edu}
\affiliation{\vs{3}Department of Physics and Astronomy, Texas A\&M University, College Station, TX 77843, USA}

\begin{abstract}
	
While quantum mechanics allows spooky action at a distance at the level of the wave-function, it also respects locality since there is no instantaneous propagation of real physical effects. We show that this feature can be proved in the standard interpretation of quantum mechanics by a simple general result involving commuting Hermitian operators corresponding to distant (causally disconnected) observables. This is reminiscent of satisfying the locality condition in relativistic quantum field theories.

\end{abstract}

\maketitle

\section{Introduction}

Quantum mechanics has a mysterious side as it is established on not so obvious and unintuative axioms like introducing a complex Hilbert space for states, Hermitian operators for observables and a rule for calculating probabilities. Although there are attempts to base its foundations to a set of reasonable principles (see e.g. \cite{ax1,ax2,ax3}), puzzling deep issues like the measurement problem are still waiting for a solution and it would not be unfair to say its fundamental structure remains to be a secret. 

Entanglement is a peculiar characteristic of quantum mechanics and it is one of the features that distinguishes quantum mechanics from a classical probabilistic theory \cite{ax3}. Moreover, entanglement is an experimentally verified feature observed in nature (see e.g. \cite{exp1,exp11,exp2,exp3,exp4}). One of the simplest example of entanglement appears in the spin-singlet state of two spin 1/2 particles $\fr{1}{\sqrt{2}}(\left|\uparrow\downarrow\right>-\left|\downarrow\uparrow\right>)$, where spins may refer to $+z$ direction. According to the standard lore (i.e. the measurement and the wave-function collapse assumptions), measuring one spin (along $+z$-direction) would {\it instantly change} the wave-function to  $\left|\uparrow\downarrow\right>$ or $\left|\downarrow\uparrow\right>$, depending on the outcome of the measurement. In that case, the other spin is {\it instantaneously determined} and yet these two particles can in principle be separated light years away from each other. 

This is clearly an example of a spooky action at a distance since it involves a sudden change possibly over great distances. The situation gets even more intriguing with the Einstein-Podolsky-Rosen (EPR) argument and Bell's theorem. One of the central themes in the EPR assertion is the concept of {\it the element of physical reality corresponding to a physical quantity,} which is related to predictions with certainty and observations without anyway disturbing a system. According to EPR in a complete theory every element of a physical reality must necessarily have a counterpart describing it. Therefore, at least in cases where the system is happen to be in an eigenstate, EPR ascribes that the wave-function should be considered as a real physical entity. Bell, on the other hand, shows a general envisaged complete theory obeying the requirements of EPR (i.e. a local hidden variable theory) contradicts with the quantum mechanical expectation values (Bell's original result involves violation of certain inequalities that arise in local hidden variable theories by quantum mechanics, but it is also possible to obtain direct contradictions in 3 and 4 particle systems \cite{bwi}).

Anyone who first learns these results and convinces himself that quantum mechanics is definitely a non-local theory is disappointed when he realizes that he cannot send {\it an instant message} using, for instance, the entangled spins. Despite the fact that quantum mechanics is non-local at the level of the wave-function, it {\it does not allow} any instantaneous propagation of {\it physical effects} (a proper analysis shows that noninteracting quantum systems cannot influence each other \cite{grf}).  

At this point one may elaborate on different definitions like strong locality, weak locality and relativistic causality (see e.g. \cite{bal}). Clearly, the relativistic causality demands additional structure to be introduced in a theory (like the speed of light or the space-time metric) and it implies locality (while the opposite is not necessarily true). In this context, attributing the physical reality of the wave-function is crucial and it mostly boils down to a philosophical preference to accept whether a theory describing the nature can be based on concepts that are not directly observable. Our aim in this paper is not to review or dwell on these deep issues. We think, despite the presence of many different views, all would agree with the minimum assertion that instantaneous propagation of a physical effect must not be allowed in a viable theory. 

It would not be surprising to see faster than speed of light evolution in non-relativistic quantum mechanics; for instance one may construct a wave-packet moving at any desired speed describing a free particle. This happens in the unitary evolution dictated by the Schr\"{o}dinger equation. In relativistic quantum field theories, the locality at the level of the unitary evolution is guaranteed by demanding the commutativity of the space-like separated field operators (which is a statement in the Heisenberg picture); for a scalar field $\f$ this becomes the condition 
\be\label{com}
\left[\phi(x^\m),\phi(y^\n)\right]=0
\ee
when $x^\m$ and $y^\n$ are space-like separated, see \cite{haag}. One should be aware of the fact that the main (seemingly obvious) problem about the violation of locality (or relativistic causality) arises in a measurement when {\it the wave-function instantaneously collapses} and not in the unitary time evolution of the states (see \cite{loc1} that discusses the locality issue in the context of quantum information processing). If the measurement and the collapse assumptions were to produce any non-local effects, this would also be a serious problem for relativistic quantum field theories. 

If we return to quantum mechanics, here we would like to show that any measurement done on a subsystem do not change the result of a complementary measurement done on the rest of the system. For example, in a general $2$-particle system, the particles can be separated from each other. We will prove that the result of the measurements done on the first particle (like measuring its spin) will not be affected in anyway from the measurements done on the second particle, even if the initial state of the system is entangled. This establishes locality since it shows that no {\it physical} spooky action at a distance is allowed when the wave-function collapses in quantum mechanics. As we will see, the commutativity of operators corresponding to distant (causally disconnected) observables, similar to \eq{com}, ensures this result. 

\section{The Canonical Example} 

Before presenting our general result, let us first discuss how locality is preserved (i.e. how instantaneous messaging is prevented) in the two spin 1/2 system prepared in the spin-singlet state
\be\label{ss1}
\fr{1}{\sqrt{2}}(\left|\uparrow\downarrow\right>-\left|\downarrow\uparrow\right>),
\ee
where we use the shorthand notation for tensor products like  $\left|\uparrow\downarrow\right>=\left|\uparrow\right>\otimes\left|\downarrow\right>$. Assume that Alice measures the first spin in the $+z$ direction and Bob measures the second one in an arbitrary direction $\hat{n}=\sin(\th)\cos(\f)\hat{i}+\sin(\th)\sin(\f)\hat{j}+\cos(\th)\hat{k}$, where $\th$ and $\f$ are the usual spherical coordinates. One can easily construct the following spin eigenstates 
\be\label{cpm}
\left|\chi_+\right>=\left[\begin{array}{c}\cos(\th/2)\\e^{i\f}\sin(\th/2)\end{array}\right],\,
\left|\chi_-\right>=\left[\begin{array}{c}\sin(\th/2)\\-e^{i\f}\cos(\th/2)\end{array}\right],
\ee
so that
\be
\vec{\s}.\hat{n}\left|\chi_\pm\right>=\pm\left|\chi_\pm\right>
\ee
where $\vec{\s}$ are Pauli spin matrices. In particular one has 
\be
\left|\uparrow\right>=\left[\begin{array}{c}1\\0\end{array}\right],\hs{3}\left|\downarrow\right>=\left[\begin{array}{c}0\\1\end{array}\right],
\ee
where these refer to spins pointing up and down in the $+z$-direction ($\th=\f=0$). 

The operators corresponding to the measurements of Alice and Bob are $\s_z\otimes I$ and $I\otimes\vec{\s}.\hat{n}$, respectively. The eigenvectors of $\s_z\otimes I$ are given by 
\be\label{eb1}
\left|\uparrow\uparrow\right>,\left|\uparrow\downarrow\right>,\left|\downarrow\downarrow\right>,\left|\downarrow\uparrow\right>
\ee
and the eigenvectors of $I\otimes\vec{\s}.\hat{n}$ can be written down as  
\be\label{eb2}
\left|\chi_+\chi_+\right>,\left|\chi_-\chi_+\right>,\left|\chi_-\chi_-\right>,\left|\chi_+\chi_-\right>. 
\ee
The spectrum of each operator is degenerate; the first two states in each set have the eigenvalue $+1$ and the last two have $-1$. 

Let us find out what Alice would observe when Bob does not make any measurements. From the expansion of the state \eq{ss1} in the eigenbasis \eq{eb1}, one finds that the probability of measuring the spin (of the first particle) to be up or down is $1/2$. Hence, Alice would observe the incoming beam to be spitted into two equal parts after passing through the Stern-Gerlach magnet.  

Assume now that Bob starts his own spin measurements (along $\hat{n}$ direction) before Alice. One can expand the state \eq{ss1} in terms of the eigenvectors \eq{eb2} so that 
\be
\fr{1}{\sqrt{2}}(\left|\uparrow\downarrow\right>-\left|\downarrow\uparrow\right>)=-\fr{e^{-i\f}}{\sqrt{2}}(\left|\chi_+\chi_-\right>-\left|\chi_-\chi_+\right>).
\ee
In obtaining the above result one may use the following identity for tensor product states:
\bea
&&\left(c_1\left|\psi_1\right>+c_2\left|\psi_2\right>\right)\otimes \left(c_3\left|\psi_3\right>+c_4\left|\psi_4\right>\right)=\nn\\
&&c_1c_3\left|\psi_1\right>\otimes\left|\psi_3\right>+c_1c_4\left|\psi_1\right>\otimes\left|\psi_4\right>+\\
&&c_2c_3\left|\psi_2\right>\otimes\left|\psi_3\right>+c_2c_4\left|\psi_2\right>\otimes\left|\psi_4\right>.\nn
\eea
Similar to the previous case, this time Bob would observe $1/2$ probability for spins up or down. 

When Bob measures the corresponding spin to be up, according to the standard rules of quantum mechanics, the wave-function (of the system)  collapses (up to an irrelevant phase) to 
\be
\left|\chi_-\chi_+\right>.
\ee
In that case the spin state of the particle observed by Alice becomes $\left|\chi_-\right>$, and thus from \eq{cpm} the probabilities of measuring spin to be up or down is given by
\be\label{p1}
P_+^{(1)}=\sin^2(\th/2),\hs{3}P_-^{(1)}=\cos^2(\th/2). 
\ee
Similarly, when Bob measures his spin to be down, the wave-function  collapses to 
\be
\left|\chi_+\chi_-\right>.
\ee
and \eq{cpm} implies 
\be\label{p2}
P_+^{(2)}=\cos^2(\th/2),\hs{3}P_-^{(2)}=\sin^2(\th/2). 
\ee
The crucial point is that {\it Bob has no control over his spin measurement.} Therefore in assessing probabilities for Alice, one should combine these possibilities statistically, e.g Alice measures his spin to be up meaning that {\it (Bob measures up and Alice measures up)} or {\it (Bob measures down and Alice measures up),} which gives
\be
P_+=\fr12 P_+^{(1)} + \fr12 P_+^{(2)}=\fr12.
\ee
Similarly one can find $P_-=1/2$. 

The upshot of the above calculation is that, according to quantum mechanics Alice observes the same probability distribution ($1/2$ for measuring the spin to be up or down) whatever Bob does, e.g. Alice cannot in anyway determine whether Bob is making a measurement or even Alice cannot ascertain if the spins she is measuring are in fact entangled. This conclusion does not undermine the special nature of the entangled state which shows itself when the measurements of Alice and Bob are combined to determine the expectation value of the operator  $\s_z\otimes \vec{\s}.\hat{n}$. In particular one has $\left<\s_z\otimes \vec{\s}.\hat{n}\right>\not=\left<\s_z\right>\left< \vec{\s}.\hat{n}\right>$, which establishes that the measurements are indeed correlated. 

\section{The general case}

Consider a system described by a wave-function $\left|\Psi\right>$, which evolves into two separate and sufficiently distant subsystems. Assume that in a given time interval locality does not allow any physical interaction to take place between the two parts and during that period Alice and Bob perform independent measurements, each person on one subsystem. Let the operators corresponding to these measurements be ${\cal A}$ and ${\cal B}$. It is not difficult to see that
\be\label{cab}
\left[{\cal A},{\cal B}\right]=0,
\ee
since ${\cal A}$ and ${\cal B}$ only act non-trivially on complementary subspaces in the Hilbert space. Indeed \eq{cab} can be adopted as a requirement for the measurements to be causally disconnected like in \eq{com}. 

For example, in a  $2$-particle system where the constituents are eventually separated from each other, Alice and Bob can setup experiments, each person making a measurement on one particle. The Hilbert space ${\cal H}$ of the system can be decomposed into the tensor product of single particle Hilbert spaces 
\be
{\cal H}={\cal H}_1\otimes {\cal H}_{1}.
\ee 
The system's wave-function $\left|\Psi\right>$ can be entangled, i.e. it can be a superposed state containing many different components in ${\cal H}_1\otimes {\cal H}_{1}$. Yet, the operators corresponding to the measurements done by Alice and Bob become ${\cal A}={\cal A}_1\otimes I$ and ${\cal B}=I\otimes {\cal B}_{1}$, which obey \eq{cab} (e.g. in the previous section ${\cal A}=\s_z\otimes I$ and ${\cal B}=I\otimes\vec{\s}.\hat{n}$). If the particles are indistinguishable, one should symmetrize or antisymmetrize the wave-function, which does not change the form of the operators. 

Once \eq{cab} is assumed, it is not difficult to show that the measurements done by Alice and Bob do not affect each others' outcomes even if the wave-function of the system instantaneously collapses. To see this one may introduce a common orthonormal eigenbasis  $\left|\psi_{ab}\right>$ in the Hilbert space obeying 
\bea
&&{\cal A}\left|\psi_{ab}\right>=\a_a\left|\psi_{ab}\right>,\nn\\
&&{\cal B}\left|\psi_{ab}\right>=\b_b\left|\psi_{ab}\right>,
\eea
and $\left<\psi_{ab}|\psi_{a'b'}\right>=\d_{aa'}\d_{bb'}$, where $\a_a$ and $\b_b$ are the eigenvalues of ${\cal A}$ and ${\cal B}$, respectively (note that ${\cal A}$ and ${\cal B}$ are commuting Hermitian operators). The state of the system can be expanded as 
\be\label{ex}
\left|\Psi\right>=\sum_{ab} c_{ab}\left|\psi_{ab}\right>.
\ee
One may take $\left|\Psi\right>$ to be normalized $\left<\Psi|\Psi\right>=1$ so that $\sum_{ab}|c_{ab}|^2=1$ and $|c_{ab}|^2$ defines a probability distribution. 

First assume that Bob does not make any measurements. In that case, the probability of Alice observing $\a_a$ can be found from the expansion \eq{ex} as
\be\label{p19}
p_a=\sum_b |c_{ab}|^2.
\ee
The eigenvalue $\a_a$ and the corresponding probability $p_a$ would completely determine the outcomes of Alice's measurements.  

Let us now assume Bob makes his measurements before Alice. Again from \eq{ex}, Bob gets the eigenvalue $\b_b$ with the probability
\be
q_b=\sum_a |c_{ab}|^2.
\ee
The eigenvalue $\b_b$ and the corresponding probability $q_b$ would completely determine what Bob gets out of his experiment.  

When Bob observes a specific eigenvalue $\b_b$, which can happen with probability $q_b$, the wave-function of the system collapses to 
\be\label{pt}
\left|\Psi\right>\to \left|\tilde{\Psi}\right>= \sqrt{\fr{1}{\sum_{a''}|c_{a''b}|^2}}\sum_{a'} c_{a'b}\left|\psi_{a'b}\right>,
\ee
where the overall constant ensures the normalization of the collapsed state $\left<\tilde{\Psi}|\tilde{\Psi}\right>=1$. Note that the denominator in \eq{pt} equals $q_b$ and one has $q_b\not=0$ since\footnote{Or else, one can simply restrict the summation over the indices corresponding to $c_{ab}\not=0$.} Bob cannot observe an outcome with $q_b=0$. When now Alice makes her measurement, the expansion of $\left|\tilde{\Psi}\right>$ in \eq{pt} implies the probability of observing $\a_a$ to be
\be
\tilde{p}_{(a|b)}=\fr{|c_{ab}|^2}{\sum_{a''}|c_{a''b}|^2}
\ee
where the notation $(a|b)$ indicates this is conditioned over Bob's observation of the eigenvalue $\b_b$. After combining\footnote{Bayes' theorem, which is an important result in statistics, can be proved by using $P(A\cap B)=P(A|B)P(B)=P(B|A)P(A)$, where $A$ and $B$ are two events, $P(A\cap B)$ is the probability of seeing $A$ and $B$ together, $P(A)$ is the unconditional probability of having $A$, similarly $P(B)$ is the unconditional probability of having $B$, $P(A|B)$ is the conditional probability of $A$ occurring once $B$ is known to occur and $P(B|A)$ is the converse. Eq. \eq{23} is a simple application of this result, where one sums over possible outcomes $B$ to find $P(A)$.} these conditional possibilities, the total probability for Alice to see the eigenvalue $\a_a$ becomes 
\be\label{23} 
p_a=\sum_b q_b\,\tilde{p}_{(a|b)}=\sum_b |c_{ab}|^2,
\ee
which is exactly equal to \eq{p19}, the value obtained with the assumption that Bob does not make any measurements. 

As a result, we see that Alice's observations are not affected at all by Bob's actions: Her outcomes are determined by the same probability distribution if Bob does not make any measurements or if Bob makes a measurement and the wave-function collapses instantaneously. There is no physical effect taking place when the wave-function collapses and quantum mechanics respect locality.

As mentioned in the Introduction, one usually envisages a problem with locality when the wave-function instantaneously collapses after a measurement and not in the unitary time evolution of the states. It is instructive to see how locality is preserved in the later case. Assuming the systems are not interacting, the Hamiltonian of the combined system can be written as $H=H_A+H_B$, where $H_A$ and $H_B$ are Alice's and Bob's Hamiltonians, respectively.  The fact that the systems are separated from each other  implies $[H_A,H_B]=0$, $[{\cal A},H_B]=0$ and $[{\cal B}, H_A]=0$. Now, Bob can change his own Hamiltonian $H_B$ but that does not affect the unitary evolution of  Alice's operator ${\cal A}$ in the Heisenberg picture. Hence Alice will get the same probability distribution for her observables, which are simply determined from the expansion of the wave-function in the eigenbasis of ${\cal A}$ that do not change in time by Bob's actions. Therefore, we see that Bob cannot affect Alice's system either if he acts on his own subsystem unitarily and the commutativity of the operators again play the main role in this case. 

\section{Conclusions} 

The problem of locality in quantum mechanics is a confusing one that can somehow lead to incorrect conclusions. As mentioned in the Introduction, the issue can be analyzed in two different settings involving either the unitary evolution dictated by the Schr\"{o}dinger equation or the wave-function collapse after a measurement. In the former case, the wave-function smoothly evolves in time and although this evolution may clash with relativistic causality in non-relativistic quantum mechanics, it does not conflict with the more broadly defined principle of locality (a non-relativistic wave may move faster than the speed of light but its evolution is still governed by a local wave-equation). In the latter case, however, there is an instantaneous change and this seems to imply a spooky action at a distance especially when entangled states are involved. 

Obviously, the instantaneous collapse of the wave-function is a non-local event but to determine whether this causes an outright  physical effect one should contemplate on how {\it physics} is defined in quantum mechanics. According to the quantum theory, the nature is inherently statistical and for any given system {\it the physics can only be attributed to the probability distributions of observables} (the task of identifying physical observables is also crucial but this is not our topic of interest now). Hence, to determine whether there is a real physical effect, one must look for a change in the probability distributions of observables. 

Once the above reasoning is accepted, it is not difficult to show that quantum mechanics is local, as we did in this work. The main crucial point is to observe \eq{cab}, which expresses that the operators corresponding to locally separated (or one may also say causally disconnected) observables commute with each other. In that case there is no change in the probability distribution of the first observable when the wave-function collapses due to a measurement done on the second one. The spooky action at a distance only takes place at the level of the wave-function and it does not imply a real physical effect. It is remarkable that the probabilistic interpretation and the collapse assumption work together beautifully to save quantum mechanics from an evident catastrophe. 

It is important to emphasize  that these results do not undermine the special nature of the entangled states. For instance, in the spin singlet state \eq{ss1}, the spin measurements of Alice and Bob are correlated with each other due to entanglement. The existence of a correlation is not surprising since the spins are coming from the same source and they were interacting with each other some time in the past, hence they cannot be treated as statistically independent random variables. However, Bell's theorem shows it is impossible to reproduce the quantum mechanical correlation by a local hidden variable theory, which makes the entanglement a purely quantum mechanical and intriguing phenomena. In any case, from the standpoint of locality the absence of instantaneous propagation of physical effects is essential and in the entangled spin system this requirement is satisfied, as we have discussed. 

To elaborate on this last point a bit more, let us imagine a {\it meta-observer} who simultaneously looks at the spin measurements of Alice and Bob along $+z$ and $-z$ directions, respectively. The  meta-observer would see a perfect anti-correlation between the spin measurements, which is not anyway surprising. However, when Alice starts rotating his magnet, our meta-observer can be amazed by the sudden loss of the perfect anti-correlation despite the great distance between Alice and Bob. The main question is  whether this phenomenon  implies non-locality or not? Our view is that this spin-spin correlation function is indeed a {\it meta-observable} and an instantaneous change of its value does not imply a spooky action of physical effects at a distance and hence  non-locality (see e.g. \cite{r1,r2} for more discussions on this issue). 

Obviously the approach presented  in this note is similar to the no-communication results like \cite{sl0, sl1}. Yet, the earlier proofs use  the language of density matrices and they do not discuss the wave-function collapse and locality. For a student who learned quantum mechanics from \cite{shk} or \cite{grfb} the presentation in this paper must be much more clearer. Moreover the main conclusion, i.e. quantum mechanics is local since the operators corresponding to distant/causally disconnected observables commute with each other, was not emphasized enough in these earlier work.

\end{document}